\documentclass[letterpaper]{jpconf}

\newcommand{\comment}[1]{\par\noindent {\em\small [#1]}}
\renewcommand{\comment}[1]{}

\newcommand{\particle}[1]{{\ensuremath{\rm #1}}}

\newcommand{\Bs}{\particle{B^0_s}}

\newcommand{\Jpsi}{\particle{J\!/\!\psi}}

\newcommand{\decay}[2]{\particle{#1\!\to #2}}

\newcommand{\BsJphi}{\decay{\Bs}{\Jpsi\phi}}             
           
\newcommand{\beq}{\begin{equation}}
\newcommand{\eeq}{\end{equation}}

\newcommand{\thetatr}{\ensuremath{\theta}}
\newcommand{\phitr}{\ensuremath{\varphi}}

\newcommand{\thetaone}{\ensuremath{\psi}}

\usepackage{graphicx}
\begin{document}
\title{Determination of the CP-violating phase in $B^0_s \to J/\psi \phi$ decays at LHCb}

\author{Greig A Cowan}

\address{University of Edinburgh\\\vspace{0.5cm}{\it on behalf of the LHCb collaboration}}

\begin{abstract}
Flavour-tagged, time-dependent, angular analysis of the decay $B^0_s \to J/\psi \phi$ allows the determination of a CP-violating phase which in the Standard Model is denoted $-2\beta_s$, and is predicted to be very small. Many models of new physics lead to significant enhancements in the value of this observable. LHCb has the capability to improve significantly the existing experimental knowledge on this phase with the data expected in the 2010 run, and to probe down to the Standard Model prediction within a few years of operation. The steps in this measurement programme will be presented. Discussion will also be given to other methods to determine this phase, and related measurements in the $B^0_s$ sector.
\end{abstract}

\section{Introduction\label{sec:jpsiphi}}

\begin{figure}[h]
	\begin{center}
	\includegraphics[scale=0.18]{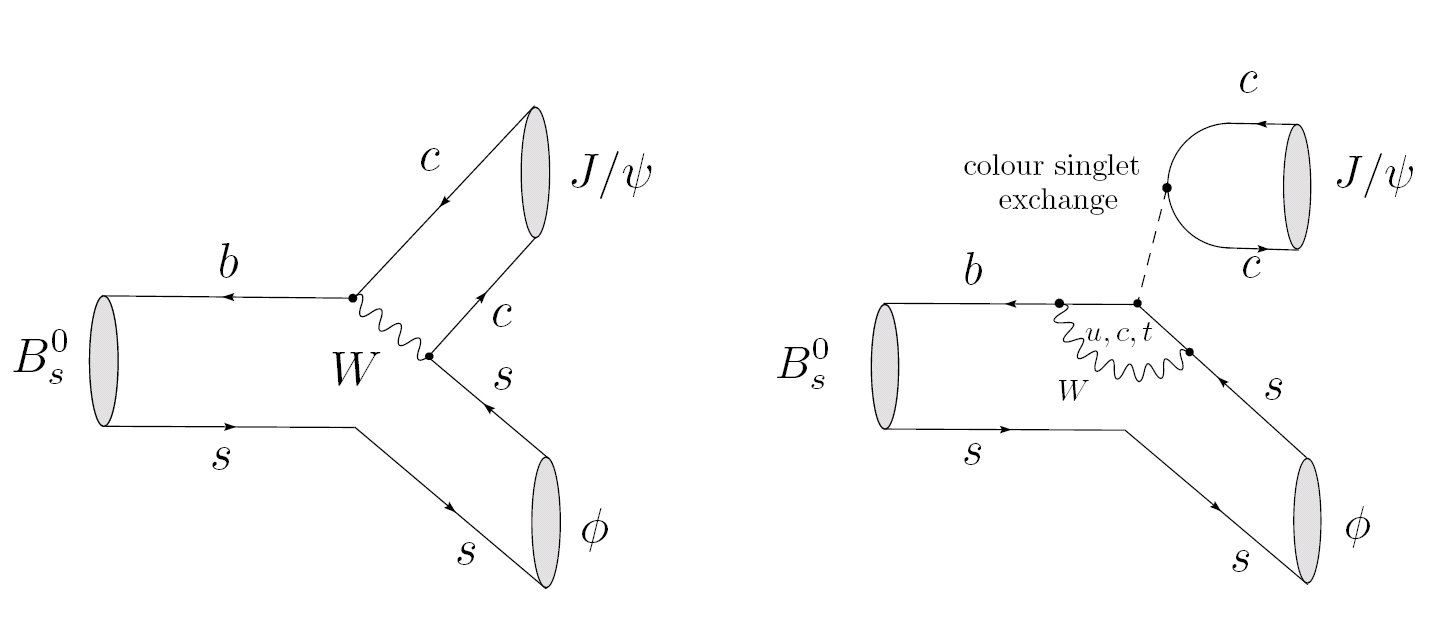}
	\end{center}
	\caption{\label{fig:feynman}Feynman diagrams contributing to the decay $B_s^0\rightarrow J/\psi\phi$ within the Standard Model. Left: tree; right: penguins.}
\end{figure}

\begin{figure}
	\vspace{-1cm}
	\begin{center}
	\includegraphics[scale=0.36]{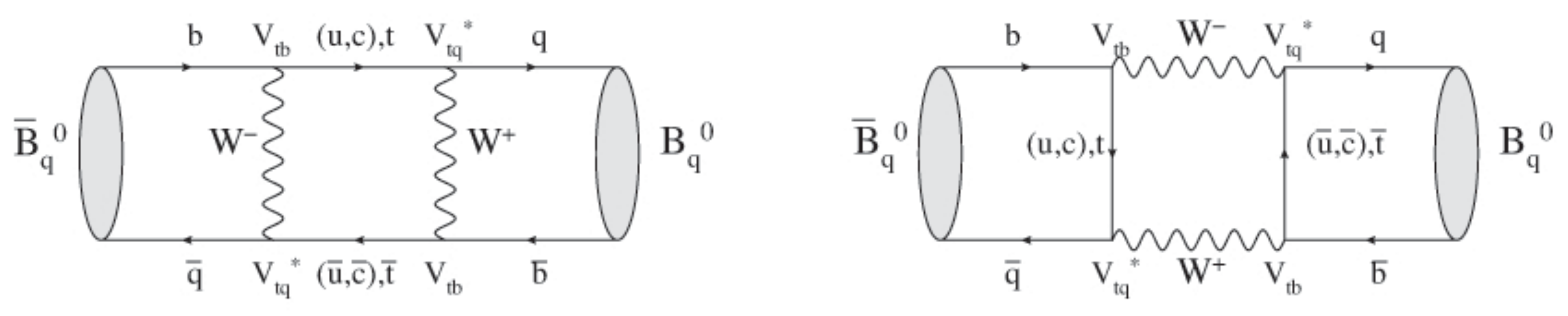}
	\end{center}
	\caption{\label{fig:mixing}Feynman diagrams responsible for $B_q-\bar{B}_q$ mixing within the Standard Model ($q=s,d$).}
\end{figure}

$B_s^0$ mesons can decay to $J/\psi\phi$ through both tree and penguin processes (figure \ref{fig:feynman}). The tree diagram dominates with a single weak phase $\Phi_D = {\rm arg}(V_{cs}V_{cb}^*)$. Before decaying, a $B_s^0$ meson can first oscillate (figure \ref{fig:mixing}) to a $\bar{B}_s^0$ meson with a mixing phase $\Phi_M = 2{\rm arg}(V_{ts}V_{tb}^*)$. Interference between mixing and decay gives  rise to a CP violating phase 
$\phi_s^{J/\psi\phi} = \Phi_M - 2\Phi_D$. In the Standard Model, this phase is predicted very
precisely  \cite{NPbetas1, NPbetas2, NPbetas3} as $\phi_s^{J/\psi\phi} = -2\beta_s + \delta P = -0.037 \pm 0.002 + \delta P$,
where $\beta_s=arg(-V_{ts}V_{tb}^*/V_{cs}V_{cb}^*)$ is the smallest angle of the ``$b-s$"
unitarity triangle. $\delta P$ is the small penguin contribution. The direct measurement of $\phi_s^{J/\psi\phi}$ is one of the key goals of the LHCb experiment at CERN. Further details about LHCb can be found in \cite{hugh, kazu}.

The box diagrams in figure \ref{fig:mixing} show the allowed oscillation of neutral $B$ mesons in the Standard Model. New particles manifesting themselves in the $B_s^0-\bar{B}_s^0$ box diagrams can significantly modify $\phi_{s,SM}^{J/\psi\phi}\to \phi_{s,SM}^{J/\psi\phi} + \phi_s^\Delta$. Recent Tevatron results give hints of deviations of $\phi_{s}^{J/\psi\phi}$ from the SM predicted value \cite{tevatron}.

\section{The $\phi_s$ analysis roadmap}

\begin{figure}
	\begin{center}
	\includegraphics[scale=0.2]{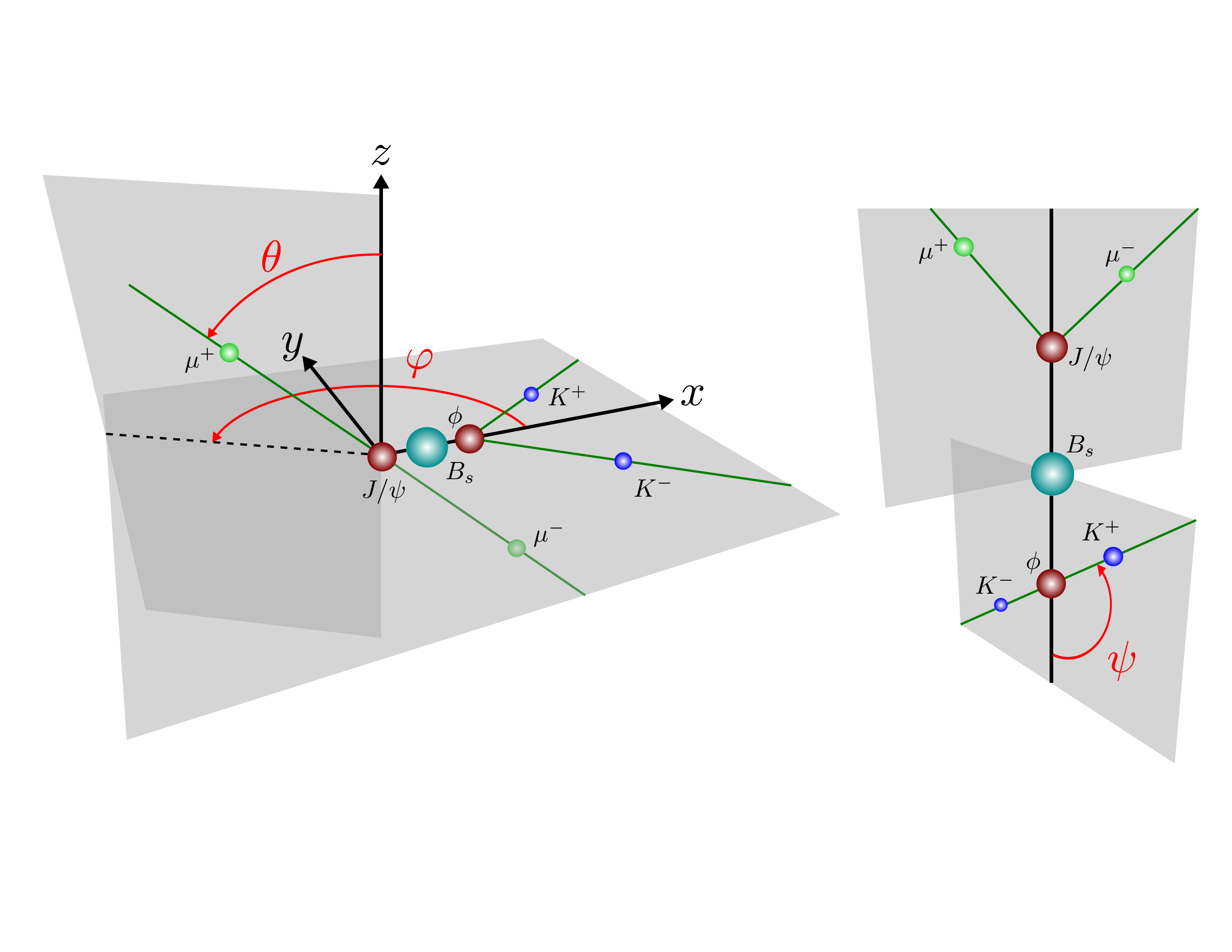}
	\end{center}
		\vspace{-1.2cm}
	\caption{\label{fig:trans_angles}Angle definition: $\theta$ is the angle formed by the positive lepton $l^+$ and the z-axis, in the $J/\psi$ rest frame. 
	The angle $\varphi$ is the azimuthal angle of $l^+$ in the same frame. In the $\phi$ meson rest frame, $\psi$ is the angle between $\vec{p}\,(K^+)$ and $-\vec{p}\,(J/\psi)$.}
\end{figure}

$B_s^0\rightarrow J/\psi(\mu\mu)\phi(KK)$ is a pseudo-scalar to vector-vector decay. Angular momentum conservation implies that the final final state is admixture of CP-odd ($\ell=1$) and CP-even ($\ell=0,2$) components where $\ell$ is the relative orbital angular momentum between the vector mesons. By performing a time-dependent angular analysis using the transversity angles $\Omega = \{\theta,\ \varphi,\ \psi\}$ (figure \ref{fig:trans_angles}) it is possible to statistically disentangle the final state.

\begin{equation}
  \frac{\partial^{4} \Gamma(\BsJphi) }{\partial t \; \partial\cos\thetatr \; \partial\phitr \;
  \partial\cos\thetaone} 
  \propto
  \sum^{6}_{k=1} h_k(t) f_k( \thetatr,\thetaone, \phitr )
  \label{eqn:diff_rate}
\end{equation}

Equation \ref{eqn:diff_rate} shows the differential decay rate for $B_s^0$ and $\bar{B_s^0}$ mesons produced as flavour eigenstates at $t=0$. The definitions of the time and angular dependent functions are given in \cite{:2009ny}. $\phi_s^{J/\psi\phi}$ typically appears in these functions multiplied terms such as $\sin(\Delta m_st)$, where $\Delta m_s$ is the mass difference of the $B_s^0$ mass eigenstates. Since these terms have opposite sign between $B_s^0$ and $\bar{B}_s^0$ the analysis benefits significantly from flavour tagging. These terms also imply that information on $\phi_s^{J/\psi\phi}$ is mostly obtained via the resolution of the fast $B_s^0$ oscillations. We extract a measurement of $\phi_s^{J/\psi\phi}$ by performing an unbinned negative log-likelihood to the propertime, $\Omega$, $B_s$ mass and initial flavour tag of the selected $B_s^0\rightarrow J/\psi\phi$ events. Additional physics parameters in the fit are the lifetime difference $\Delta \Gamma_s$ between the mass eigenstates, the average lifetime $\bar{\Gamma}_s$, the CP amplitudes ($A_\perp, A_\parallel, A_0$) and corresponding strong phases ($\delta_\perp, \delta_\parallel, \delta_0$). Parameters describing the background, LHCb detector propertime resolution and mistag efficiency ($\omega$) and angular acceptance corrections are included in equation \ref{eqn:diff_rate}.

Dedicated LHCb Monte Carlo simulation predicts a $B_s^0\rightarrow J/\psi\phi$ signal yield of 117k events in $2fb^{-1}$ of data, with a total trigger efficiency of $\sim 70\%$. The trigger and selection is defined so as to minimise any bias on the reconstructed propertime and angular observables while maximising sensitivity to $\phi_s^{J/\psi\phi}$. There are two main backgrounds in this analysis: prompt and long-lived. The prompt component is mainly produced by combinations of tracks from the primary vertex and has a very small lifetime. The long-lived background has at least one track coming from a secondary vertex from $b-$ or $c-$hadron decay. The $B/S$ in a $50MeV/c^2$ mass window around the $B_s$ meson mass is 1.6 for the prompt and 0.5 for the long-lived background respectively.

The main sources of systematic uncertainty in this measurement come from the geometrical shape of LHCb giving rise to a relative 5\% variation in the angular observables; the flavour tagging ($\omega = 34\pm1\%$) and the propertime resolution ($38\pm5fs$). Each of these results in a relative bias on $\phi_s^{J/\psi\phi}$ of $\sim 7\%$ \cite{:2009ny}. It has also been shown \cite{Xie:2009fs} that ignoring a possible 5-10\% S-wave contamination in the $\phi\to KK$ decay would lead to a 15\% bias on the measured value of $\phi_s^{J/\psi\phi}$ while including it in the fit will reduce the sensitivity by around 20\% .

\section{LHCb sensitivity to $\phi_s$}

\begin{figure}
	\vspace{-1.5cm}
	\begin{center}
	\includegraphics[scale=0.34]{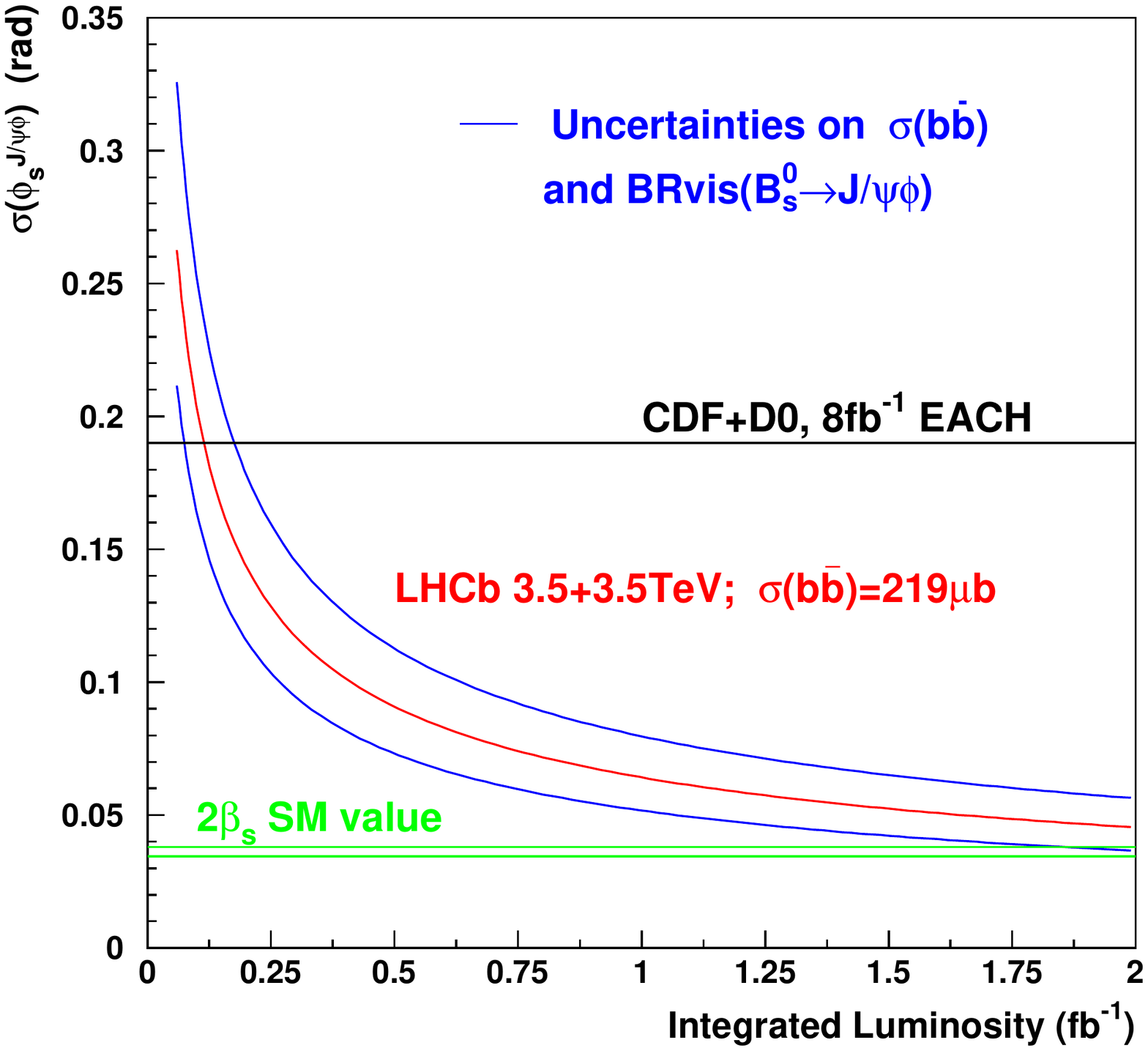}
	\includegraphics[scale=0.34]{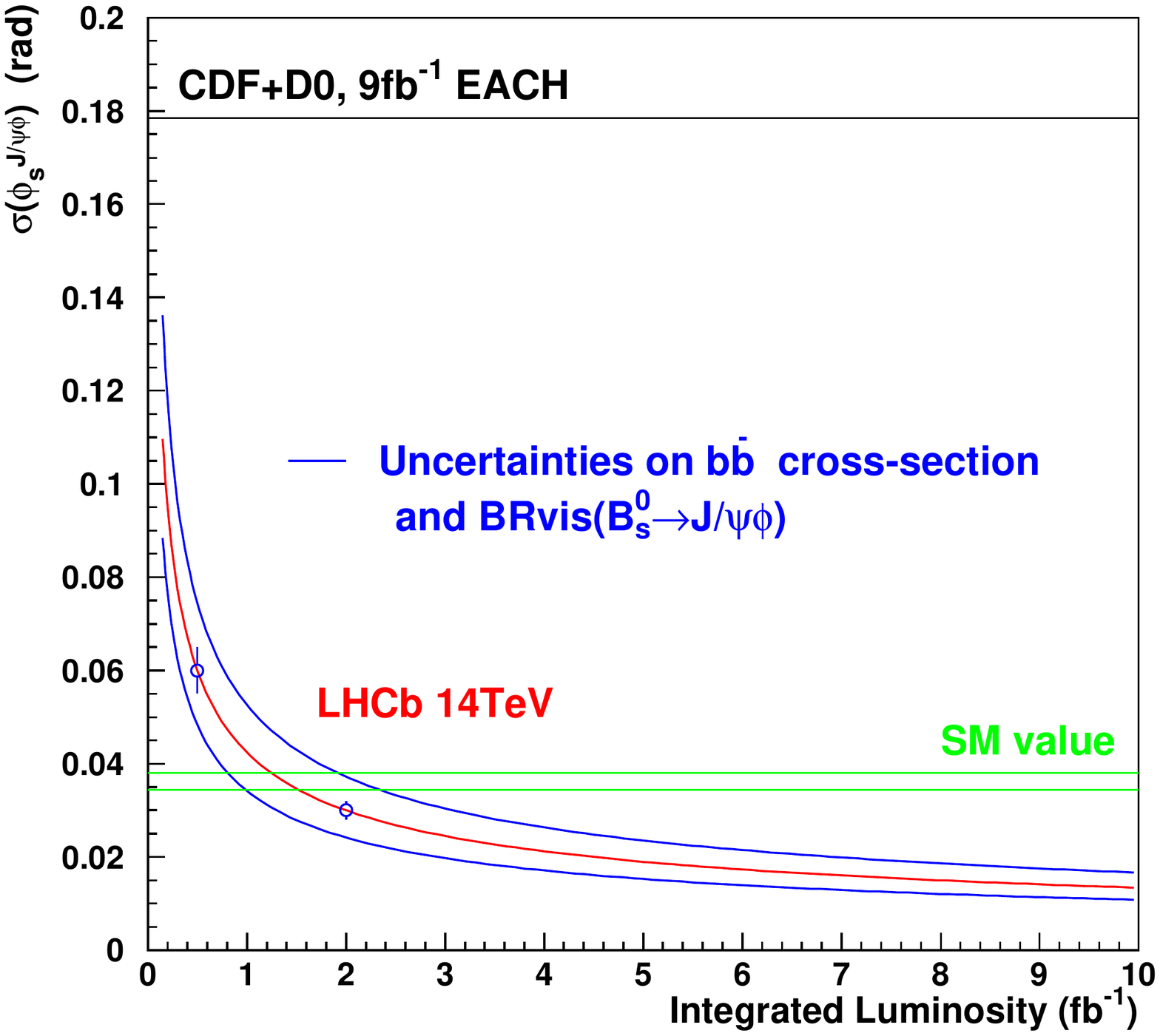}
	\end{center}
	\vspace{-1.5cm}
\caption{\label{fig:sensitivity}Red line: Statistical uncertainty on $\phi_s$ versus the integrated luminosity. Blue
line: Uncertainties coming from the bb cross-section and the visible branching ratio on $B_s^0\rightarrow J/\psi\phi$. The green band is the Standard Model value: $2\beta_s = (0.0360^{+0.0020}_{-0.0016})$ rad \cite{NPbetas1}. Left: from 0 to $2fb^{-1}$, assuming a centre-of-mass energy of 7TeV. Right: from 0 and $10fb^{-1}$, for proton-proton collisions at the LHC design centre-of-mass energy of 14TeV.
	}
\end{figure}

Using many hundreds of toy Monte Carlo experiments, the statistical uncertainty on $\phi_s^{J/\psi\phi}$ has been estimated for different integrated luminosities and LHC centre-of-mass collision energies. Figure \ref{fig:sensitivity} shows the statistical precision on the measurement of $\phi_s^{J/\psi\phi}$ as a function of integrated luminosity. During 2010-2011 7TeV run of the LHC, if $\phi_s^{J/\psi\phi}\sim 0.7$ rad then LHCb can discover New Physics in $\bar{B}^0_s-B^0_s$ mixing (i.e., establish $\phi_s^{J/\psi\phi}\neq -2\beta_s$ at $5\sigma$) with $0.3 fb^{-1}$ of data. Under nominal running conditions of 14TeV then after collecting $2fb^{-1}$, corresponding to 117k events after trigger, the LHCb sensitivity to $\phi_s^{J/\psi\phi}$ will be $0.03$ rad.

\section{Alternative routes to $\phi_s$}

\begin{figure}
	\vspace{-1.2cm}
\begin{center}
\begin{tabular}{cc}
\begin{minipage}{0.4\linewidth}
	\includegraphics[scale=0.35]{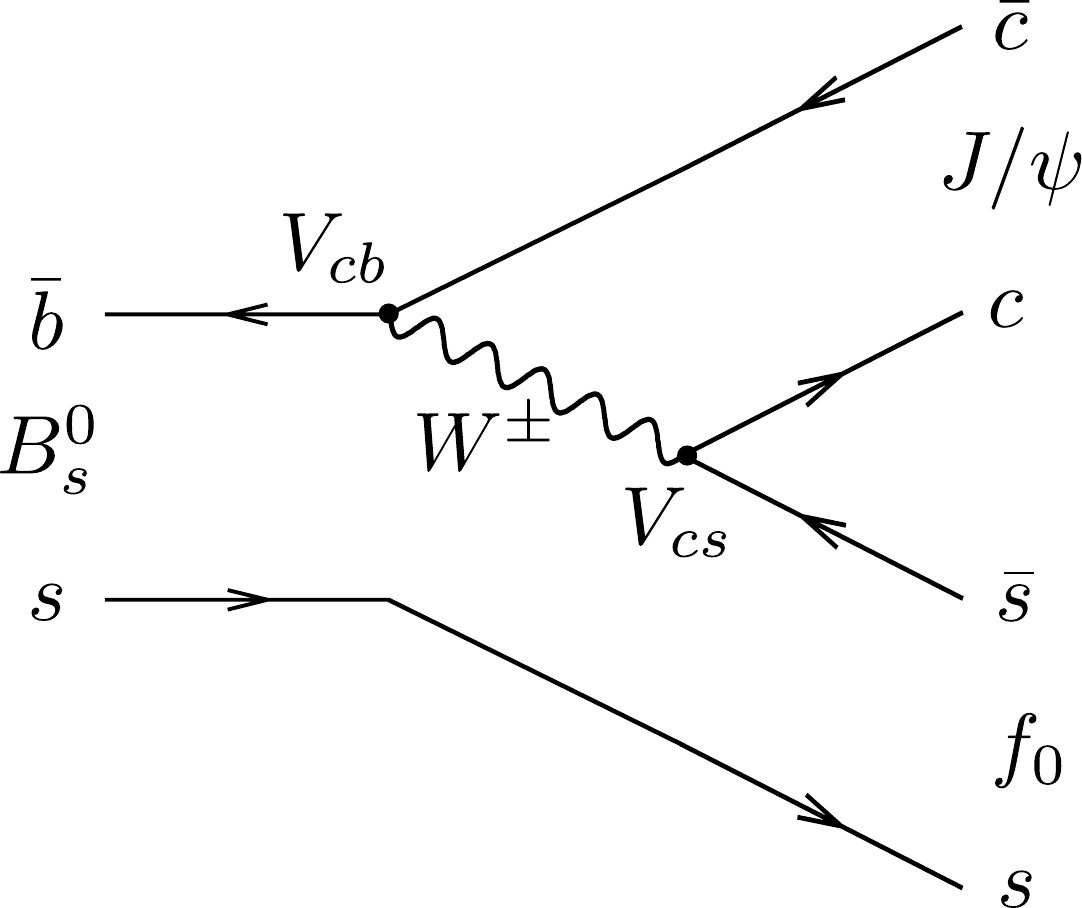}
\end{minipage}
&
\begin{minipage}{0.3\linewidth}
	\includegraphics[scale=0.3]{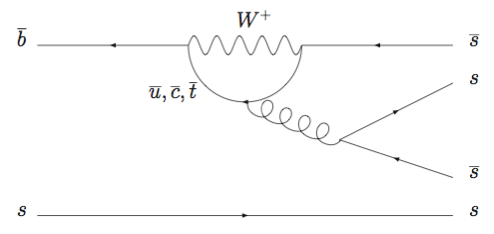}
\end{minipage}
\end{tabular}
	\caption{\label{fig:f0} Feynman diagrams responsible for $B_s^0 \rightarrow J/\psi f_0$ (left) and $B_s^0 \rightarrow \phi\phi$ (right) within the Standard Model.}
\end{center}
\end{figure}

In addition to $B_s^0\to J/\psi \phi$, modes such as  $B_s^0\to J/\psi\eta$ and $B_s^0\to J/\psi\eta^\prime$ can provide access to the same CP violating phase, $\phi_s$. However, these modes are pure CP eigenstates, implying that no complex angular analysis is required in order to disentangle the final states. Since $\eta, \eta^\prime$ decay modes contain more than one photon, they typically have lower reconstruction efficiency than modes decaying to charged particles. This makes $B_s^0\to J/\psi f_0(980)$ where $f_0\to\pi^+\pi^-$ (figure \ref{fig:f0}, left) an interesting decay to study \cite{Stone:2009hd}.

With $2fb^{-1}$ LHCb expects to collect 26.1k $B_s^0\to J/\psi f_0(980)$ events, leading to a sensitivity on $\phi_s$ of $0.044$ rad. This is larger than the sensitivity from $B_s^0\to J/\psi \phi$ alone, but this analysis will provide an important cross check on the vector-vector result.

\section{CP measurements using penguin decays}

$B_s^0\rightarrow \phi\phi$ is a flavour changing neutral current process, forbidden at tree level in the Standard Model and as such, the gluonic $b\rightarrow s$ penguin (figure \ref{fig:f0}, right) dominates the decay with a phase $\Phi_D = 2{\rm arg}(V_{ts}V_{tb}^*)$. This decay has access to the same weak mixing phase $\Phi_M$ as defined in section \ref{sec:jpsiphi} implying that in the Standard Model the CP violating phase $\phi_s^{\phi\phi} = \Phi_M-\Phi_D = 0$. However, non-Standard Model particles could manifest themselves in both the $B_s^0$ mixing and penguin decay, giving rise to new phases which would cause $\phi_s^{\phi\phi}$ to deviate from zero. Therefore, any measurement of $\phi_s^{\phi\phi}\neq 0$ would imply the existence of New Physics.

Like $B_s^0 \rightarrow J/\psi\phi$, $B_s^0\rightarrow \phi(KK)\phi(KK)$ is a pseudoscalar to vector-vector decay. We can perform a similar angular analysis in order to disentangle the CP components in the final state. With $2fb^{-1}$, LHCb expects to collect 6.2k events leading to an expected LHCb sensitivity to $\phi_s^{\phi\phi}$ of $0.14$ rad. The background over signal ratio is estimated from a sample of inclusive bb to be below 0.8 at 90\% CL \cite{Styles}.

\section{Summary}

The channels $B_s^0\rightarrow J/\psi\phi$,  $B_s^0\rightarrow J/\psi f_0$ and  $B_s^0\rightarrow \phi\phi$ will allow LHCb to probe New Physics contributions in $B_s$ mixing and $b\rightarrow s$ penguin decays. I have outlined the analysis strategy that LHCb will follow to measure the CP violating phases in these modes. With $2fb^{-1}$ of data LHCb will deliver a sensitivity on $\phi_s^{J/\psi\phi}$ of $0.03$ rad and on $\phi_s^{\phi\phi}$ of $0.14$ rad.\\

\end{document}